\documentclass{article}
\usepackage{graphics}
\title{Trapped surfaces in prolate collapse in the Gibbons-Penrose
construction}
\author{M.A. Pelath\thanks{Enrico Fermi Institute and Department of Physics,
University of Chicago, 5640 S. Ellis Avenue, Chicago, Illinois 60637-1433},
K.P. Tod\thanks{Mathematical Institute and St John's College, Oxford OX1 3LB, 
UK}, and Robert M. Wald\thanks{Enrico Fermi Institute and Department of 
Physics,
University of Chicago, 5640 S. Ellis Avenue, Chicago, Illinois 60637-1433}}
\date{\today}
\begin{document}
\maketitle
\begin{abstract}
We investigate existence and properties of trapped surfaces in two
models of collapsing null dust shells within the Gibbons-Penrose
construction. In the first model, the shell is initially a prolate
spheroid, and the resulting singularity forms at the ends first
(relative to a natural time slicing by flat hyperplanes), in analogy
with behavior found in certain prolate collapse examples considered by
Shapiro and Teukolsky. We give an explicit example in which trapped
surfaces are present on the shell, but none exist prior to the last
flat slice, thereby explicitly showing that the absence of trapped
surfaces on a particular, natural slicing does not imply an absence of
trapped surfaces in the spacetime. We then examine a model considered
by Barrab\`{e}s, Israel and Letelier (BIL) of a cylindrical shell of
mass \(M\) and length \(L\), with hemispherical endcaps of mass
\(m\). We obtain a ``phase diagram'' for the presence of
trapped surfaces on the shell with respect to essential
parameters \(\lambda \equiv M/L \) and \(\mu \equiv m/M \). It is found
that no trapped surfaces are present on the shell
when \(\lambda\) or \(\mu\) are sufficiently small.  (We
are able only to search for trapped surfaces lying on the shell
itself.) In the limit \(\lambda \rightarrow 0\), the existence or
nonexistence of trapped surfaces lying within the shell is seen to be
in remarkably good accord with the hoop conjecture.
\end{abstract}

\section{Introduction}
It is well known from the singularity theorems \cite{he:1973} that
singularities must occur in a wide variety of circumstances relevant
to gravitational collapse---in particular, when trapped surfaces or
reconverging light cones \cite{rp:1998} are present. However, there is
still little direct evidence on whether cosmic censorship
holds, \emph{i.e.}, whether the singularities of gravitational collapse are
always hidden in black holes (see \cite{rp:1998} and \cite{rw:1998}
for recent reviews). Much of our knowledge of what occurs in
gravitational collapse is based upon the analysis of spherically
symmetric models and their linearized perturbations. Without results
from other models, we cannot know which features might be shared
by---and which absent from---general kinds of collapse.

It has been suggested by a number of authors that \emph{highly prolate} 
gravitational collapse may be of a qualitatively different character
from spherical collapse (see, \emph{e.g.\ }\cite{rp:1969}). An explicit 
statement of this view has been expressed in the ``hoop conjecture'' 
\cite{kt:magic}:
\begin{quote}
Horizons form when and only when a mass \(M\) gets compacted into a region
whose circumference in EVERY direction is \(C \stackrel{<}{\sim} 4 \pi M\).
\end{quote}
The intended interpretation of this statement is that a distribution
of matter that is elongated in one dimension but is sufficiently
narrow in the other dimensions might be expected to collapse without
the creation of a event horizon, so that any singularity that formed
would be ``naked''.  However, there are serious difficulties in giving
a precise statement of the hoop conjecture, both in defining the
``mass''---one would want to include concentrations of gravitational
energy, for which there is no local definition---and the
``circumference''---which always can be made arbitrarily small if no choice
of spacelike slicing is specified.

Nevertheless, numerical simulations by Shapiro and Teukolsky
\cite{st:1991} of the collapse of a collisionless gas cloud appeared
to lend support to the hoop conjecture. For a sufficiently prolate
cloud, a singularity was found to form (after which point the
numerical evolution could not be continued) but no trapped surfaces
were found prior to the singularity on the family of maximal
hypersurfaces they used for their time evolution.

Barrab\`{e}s, Israel and Letelier \cite{bil} considered a simple
analytic counterpart of the Shapiro-Teukolsky examples within the
Gibbons-Penrose construction (see below). This model consists of
an imploding shell of photons (null dust) in the shape of a cylinder
with hemispherical endcaps and with piecewise uniform mass-energy
density. They found, similarly, that given a fixed amount of matter,
if the cylindrical portion of the shell is made long enough 
(\emph{i.e.\ }sufficiently prolate), the 2-surfaces formed by the 
intersection of the shell with flat hyperplanes of the Minkowski 
region interior to the shell are never trapped at any stage of
collapse.

However, in both the numerical and analytic models, the conclusion
that no trapped surfaces exist relies on a specific choice of time
slicing.  Indeed, as a counterargument to the Shapiro-Teukolsky
result, Wald and Iyer \cite{wi:1991} proved that even in
Schwarzschild spacetime, it is possible to choose a (highly
non-spherical) time slice which comes arbitrarily close to the
singularity, yet for which no trapped surfaces are found to its
past. This example explicitly demonstrates that the absence of trapped
surfaces in a particular time slicing does not establish the absence of
trapped surfaces in the spacetime. Nevertheless, the slicing in
\cite{wi:1991} is rather contrived, and it is far from clear that
anything similar to this is occurring in the Shapiro-Teukolsky
examples.

The main purpose of this paper is to investigate the existence and
properties of trapped surfaces in highly prolate collapse in the
context of spacetime models produced by the Gibbons-Penrose
construction. The Gibbons-Penrose spacetimes contain imploding shells
of null dust, with the region of spacetime interior to the shell being
flat. These models have the advantage that it is relatively easy to
search for trapped surfaces lying on (or, more precisely, infinitesimally 
outside of) the collapsing shell. Unfortunately, in these models, the spacetime
outside of the shell is not explicitly known, so one cannot determine
the existence or properties of trapped surfaces lying at a finite
distance outside of the shell.

In section 2, we review the Gibbons-Penrose construction.  In section
3, we examine the collapse of an initially prolate spheroidal
shell. We obtain an explicit example---previously suggested in
\cite{kpt:1992}---in which, in the natural flat hyperplane time
slicing in the region interior to the shell, the singularity forms at
the poles before any trapped surface has finished forming along the
equator. Thus, a numerical code using this natural time
coordinate---which would halt at the time the singularity begins to
form---would not find any trapped surfaces on the shell.
Nevertheless, in the fully evolved spacetime, trapped surfaces do
occur on the shell. Thus, this example explicitly displays the
behavior closely analogous to that constructed in \cite{wi:1991}, but
does so in the context of prolate collapse with a natural choice of
time slicing.

In section 4, we re-examine the Barrab\`{e}s-Israel-Letelier (BIL)
model of a collapsing cylindrical shell with hemispherical endcaps, in
the case where \(\lambda = M/L < 1/8\), where \(M\) is the mass of the
cylinder and \(L\) its length. BIL showed that in this case, none of the
2-surfaces formed by the intersection of the shell with the flat
hyperplanes of the interior region are trapped surfaces. Here, we
search for trapped surfaces on the shell which do not lie in these
hyperplanes. We present a ``phase diagram'' for the the presence or
absence of trapped surfaces as a function of the essential
scale-invariant parameters \(\lambda \) and \(\mu \equiv m/M \), where
\(m\) the mass of each endcap. We find that no trapped surfaces are
present on the shell for \(\lambda\) or \(\mu\) sufficiently small
(although we expect that trapped surfaces will be present outside of
the shell in all cases). We find that the existence or nonexistence
of trapped surfaces lying on the shell is in remarkably good agreement
with what would be expected from the hoop conjecture.

\section{The Gibbons-Penrose construction}

The Gibbons-Penrose construction considers a thin shell of 
matter collapsing inward at the speed of light from past null infinity. The 
trajectory of the shell describes a null hypersurface \(\mathcal{N}\) which 
divides the spacetime into two regions: the interior, which is Minkowski, and 
the exterior, which is not flat and generally contains gravitational radiation.
Singularities form where the null hypersurface develops caustics.

The construction is completely specified by two functions, one defining the
shape of the collapsing shell, and one defining its (distributional) mass
density. The degrees of freedom associated with the shape correspond to
a choice of 2-surface \(\mathcal{S}(T)\) in the hyperplane \(t=T\)
of Minkowski spacetime. The hypersurface \(\mathcal{N}\) is then generated by
the ingoing normal null geodesics through \(\mathcal{S}(T)\). We restrict 
to the case where \(\mathcal{S}(T)\) is convex, so that no caustics appear 
to the past of our initial time. The ``initial'' mass density \(\sigma(T)\) 
can be chosen to be any non-negative function on \(\mathcal{S}(T)\).
The mass density is then extended to the whole of \(\mathcal{N}\) by the 
conservation equation. 

One can then, in principle, go about searching for trapped surfaces in
the spacetime. In general, this, of course, requires solving Einstein's 
equations in the region of interest. The great advantage of the 
Gibbons-Penrose construction is that as long as one focuses attention on 
surfaces infinitesimally outside \(\mathcal{N}\), one need not know anything 
about the spacetime outside \(\mathcal{N}\); that is,
given some candidate compact 2-surface in \(\mathcal{N}\), we can determine
whether it is trapped without ever solving for the exterior spacetime.
The expansion \(\theta\) of future-directed, outgoing null geodesics normal 
to the 2-surface, just before they cross the shell, is \cite{gg:1997}
\begin{equation}
\theta_{\mathrm{inside}} = - \mbox{Tr}~L
\end{equation}
where \(L\) is the extrinsic curvature of the 2-surface under its embedding
in Minkowksi spacetime, with respect to outgoing null normals. Just outside 
the shell, we have
\begin{equation}
\theta_{\mathrm{outside}} = \theta_{\mathrm{inside}} + 
    (\Delta \theta)_{\mathrm{matter}}
\end{equation}
where, by integrating the Raychaudhuri equation \cite{rw:gr} across the
shell (so that only the distributional source term contributes), we
find that
\begin{equation}
(\Delta \theta)_{\mathrm{matter}} = -8 \pi \sigma
\end{equation}
Thus, the 2-surface is trapped if
\begin{equation}
-\mbox{Tr}~L < 8 \pi \sigma
\end{equation}
everywhere on the surface.

The project is further simplified if one looks instead for an 
\emph{outer marginally-trapped surface} (MTS), \emph{i.e.\ }a compact 
2-surface for which the expansion of future-directed outgoing null normals is 
everywhere zero. By an argument similar to one given in \cite{he:1973} for 
trapped surfaces lying in a spacelike hypersurface, if \(\mathcal{N}\) 
contains trapped surfaces, then the outer boundary of the region containing
the trapped surfaces is an MTS. The absence of an outer 
marginally-trapped surface on \(\mathcal{N}\) thus implies the absence of 
trapped surfaces on \(\mathcal{N}\).

The advantage of searching for MTS's is that a MTS satisfies the equation
\begin{equation} 
-\mbox{Tr}~L = 8 \pi \sigma 
\label{MTS}
\end{equation}
rather than an inequality. Thus, rather than attempt to explore the myriad
of possibilities for where trapped surfaces might occur, one studies eq.\ 
(\ref{MTS}) and attempts to determine if any global solutions exist.

In the following, we narrow our search for trapped surfaces in the
whole spacetime to the search for MTS's confined to the null hypersurface
\(\mathcal{N}\).

\section{The collapsing spheroid}

We now consider the collapse of a prolate ellipsoid in the Gibbons-Penrose
construction. We choose Minkowski coordinates \((t,x,y,z)\) in what will
be the interior of \(\mathcal{N}\), and specify the initial shape 
\(\mathcal{S}_{0}\) at \(t=0\) to be an axisymmetric ellipsoid (\emph{i.e.},
a spheroid)
\begin{equation}
x^{\mu}(\theta,\phi) = (0,a \sin \theta \cos \phi,a \sin \theta \sin \phi,
  b \cos \theta)
\end{equation}
\begin{equation}
\theta \in [0,\pi], \phi \in [0,2 \pi)
\end{equation}
where \(a<b\) so that the spheroid is prolate. Following the ingoing null 
normals forward and backward 
in time yields the form of \(\mathcal{N}\),
\begin{equation}
x^{\mu}(t,\theta,\phi) = 
  \left( t, \left( a-\frac{bt}{\sqrt{E}} \right) \sin \theta \cos \phi,
            \left( a-\frac{bt}{\sqrt{E}} \right) \sin \theta \sin \phi, 
            \left( b-\frac{at}{\sqrt{E}} \right) \cos \theta \right)
\label{formofN}
\end{equation}
where
\begin{equation}
E(\theta) = a^{2} \cos^{2} \theta + b^{2} \sin^{2} \theta
\end{equation}

The metric induced on the intersection of \(\mathcal{N}\) with the hyperplanes 
of constant \(t\) is
\begin{equation}
ds^{2} = \left(1-\frac{abt}{E^{3/2}}\right)^{2} E d\theta^{2} + a^{2}
     \left(1-\frac{bt}{aE^{1/2}}\right)^{2} \sin^{2} \theta d\phi^{2}
\label{inducedmetric}
\end{equation}
This metric becomes truly singular when the quantities in parentheses
become zero. This is just the condition that \(t\) is equal to a principal 
radius of curvature of the initial spheroid, where these radii of curvature
are given by
\begin{equation}
R_{1}^{-1} = \kappa_{1}(\theta) = \frac{ab}{E^{3/2}}
\end{equation}
\begin{equation}
R_{2}^{-1} = \kappa_{2}(\theta) = \frac{b}{a E^{1/2}}
\end{equation}
Since \(\kappa_{1} \leq \kappa_{2}\) (with equality only at the poles), the
singularity actually occurs at \(t(\theta)=1/\kappa_{2}(\theta)\).

The evolution of the initial mass density \(\sigma_{0}\) (which so far is 
unspecified) is governed by the local mass-conservation equation
\begin{equation}\frac{\partial}{\partial t} (\sigma dA)=0\end{equation}
where \(dA\) is the area form obtained from the metric
(\ref{inducedmetric}),
\begin{equation}
dA = a U V \sqrt{E} \sin \theta~ d\theta \wedge d\phi
\end{equation}
where
\begin{equation}
U(\theta,t) = 1 - \kappa_{2}t
\end{equation}
\begin{equation}
V(\theta,t) = 1 - \kappa_{1}t
\end{equation}
Given some \(\sigma_{0}\), then,
\begin{equation}
\sigma(t,\theta,\phi) = \frac{\sigma_{0}}{UV}
\label{conservation}
\end{equation}
Equations (\ref{formofN}) and (\ref{conservation}) define the Gibbons-Penrose
construction for the collapsing spheroid. 

Now consider an arbitrary spacelike 2-surface on \(\mathcal{N}\),
\begin{equation}
t=F(\theta,\phi) 
\label{Fdef} 
\end{equation}
We seek the condition on \(F\) such that this surface is marginally-trapped.
The embedding of the 2-surface in Minkowski spacetime is given by equations
(\ref{Fdef}) and (\ref{formofN}), and its extrinsic curvature is readily
calculated. Combining this result with (\ref{conservation}), we find that
eq.\ (\ref{MTS}) becomes
\begin{equation}
 \nabla^{2}F + \frac{1}{2} \left( \frac{\kappa_{1}}{V} + 
  \frac{\kappa_{2}}{U} \right)
  (|\nabla F|^{2} + 1) = 8 \pi \frac{\sigma_{0}}{UV}
\label{yuck}
\end{equation}
where Laplacians and gradients are with respect to the metric 
(\ref{inducedmetric}) and the functions \(U\) and \(V\) are to be evaluated on 
the 2-surface (\emph{i.e.\ }replace \(t\) by \(F(\theta,\phi)\) everywhere).

Finally, we will restrict attention to axisymmetric solutions \(F=F(\theta)\)
since the outer boundary of the region containing trapped surfaces must have
the same symmetries as the spacetime itself. In this
case eq.\ (\ref{yuck}) reduces to 
\begin{eqnarray}
\frac{U}{V \sqrt{E} \sin \theta} \frac{d}{d \theta}
  \left( \frac{\sin \theta}{\sqrt{E}} F' \right)
  + (3 \kappa_{1} - \kappa_{2} - 2 \kappa_{1} \kappa_{2} F)
    \frac{F F'}{2 E V^{2}}
    \left( \frac{F'}{F} - \frac{E'}{E} \right) \nonumber \\
  + \frac{1}{2} (\kappa_{1} + \kappa_{2} - 2 \kappa_{1} \kappa_{2} F)
  = 8 \pi \sigma_{0}
\label{MTSequation}
\end{eqnarray}

\begin{figure}
\centering
\includegraphics{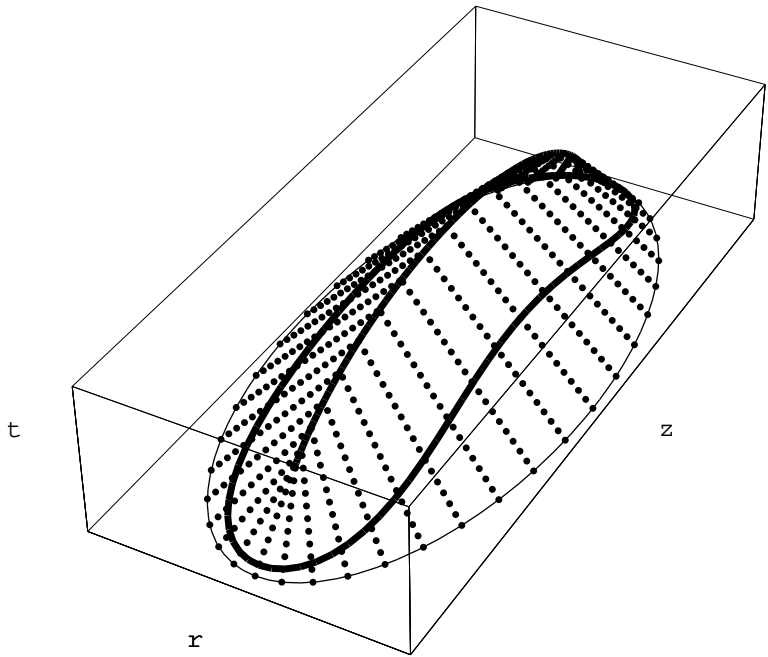}
\caption{The chosen MTS (\(\gamma=1/2,~a=1,~b=\sqrt{5}\)) and the singularity 
(\(\phi\)-coordinate suppressed). Dotted lines indicate null generators of 
the hypersurface \(\mathcal{N}\). Note that the MTS forms later at the 
equator than the singularity forms at the poles.}
\label{3DMTS}
\end{figure}

The parameters and free functions of this model can be made to yield a 
spacetime where the outermost marginally-trapped surface encloses the 
singularity, but forms later at the equator than the singularity does at the
poles, in the time coordinate \(t\). A numerical code using this time 
coordinate would then halt before any trapped surfaces have formed. 
To obtain an explicit example where this occurs, we will specify a
form of \(F\), and use (\ref{MTSequation}) to determine the requisite initial 
mass density, which we must verify to be everywhere positive and finite. 
A relatively simple choice of MTS is the surface of constant \(V\) 
\begin{equation}
F(\theta) = \frac{\gamma a^{2}}{b^{2} \kappa_{1}}
\end{equation}
where \(\gamma\) is some constant. In order that the MTS have the special
properties described above, we must restrict \(\gamma\) according to
\begin{equation}\frac{a}{b}<\gamma<1\end{equation}
Substituting \(F\) into (\ref{MTSequation}) gives the initial mass density,
\begin{equation}\sigma_{0}=\frac{\kappa_{1}}{8 \pi} (A \cos^{6} \theta + 
 B \cos^{4} \theta + C \cos^{2} \theta + D)\end{equation} 
where \(A, B, C\) and \(D\) are constants depending on \(\gamma\), \(a\) and 
\(b\). For the choice 
of parameters \(\gamma=1/2,~a=1,~b=\sqrt{5}\), for example,
we find \(A=-352/225,~B=8/9,~C=2/9,~\mbox{and}~D=11/6\), from which 
it can be verified that \(\sigma_{0}\) is everywhere positive. The MTS
for this case is plotted in fig. \ref{3DMTS}.

The possibility exists, however, that our chosen MTS may not be the
outermost one, and so would not be the outer boundary of a trapped region.
(Examples where there exists an MTS which is not the outer boundary of
a trapped region will be given in the next section.) In that case trapped
surfaces would occur to the past of our MTS, and our argument about the
inability to find trapped surfaces would not apply. We have investigated
this issue numerically and have found that for at least some 
choices of parameters \(a\), \(b\), and \(\gamma\), our MTS is indeed the
outermost one.

We numerically examined the boundary value problem consisting of 
(\ref{MTSequation}) together with \(F'(\pi/2)=0\) (to ensure reflection 
symmetry about the \(\theta=\pi/2\) plane) and \(F'(0)=0\) (to ensure 
regularity at the poles). The singular nature of the differential equation at
\(\theta=0\) prevents the 
direct application of the boundary conditions in the numerical routines. 
Instead, the equation is solved from \(\theta=\pi/2\) toward \(\theta=0\), 
with initial conditions \(F'(\pi/2)=0\) and \(F(\pi/2)=t_{0}\). 
We then check to see if the boundary condition \(F'(0)=0\) is met. 

For the choice of parameters \(\gamma=1/2,~a=1,~b=\sqrt{5}\), the
chosen MTS has \(t_{0}=0.5\). For \(t_{0}\) slightly less than 0.5,
the solutions ``curve more to the past'' and \(F'(\theta) \rightarrow
\infty~\mbox{as}~\theta \rightarrow 0\). As \(t_{0}\) is further 
decreased, the solutions diverge even more rapidly as 
\(\theta \rightarrow 0\), and the boundary condition \(F'(0)=0\) cannot be 
met. We conclude 
that the chosen MTS is the outermost one, and therefore no trapped 
surfaces exist outside of our chosen MTS.\footnote{For \(t_{0}\) slightly 
larger than 0.5, the solutions ``curve more toward the future'' and 
\(F'(\theta) \rightarrow -\infty~\mbox{as}~\theta \rightarrow 0\);
these surfaces can be perturbed slightly to construct regular trapped 
surfaces bounded by our chosen MTS. When \(t_{0}\) is slightly 
larger than 0.51, solutions run into the singularity before reaching
\(\theta=0\). We therefore also conclude that there is no MTS forming
at a later time than our chosen MTS.}

\section{The Barrab\`{e}s-Israel-Letelier model}

In the Barrab\`{e}s-Israel-Letelier model (BIL) model of prolate collapse,
the initial shape is a cylinder of length \(L\) with hemispherical endcaps.
As this shell collapses, it retains this form; this is one advantage of
this model. The resulting singularity will be a ``spindle'' singularity
with stronger, point singularities at the ends. 

Again, we let \((t,x,y,z)\) be Minkowski coordinates for the region interior
to the shell. Since the singularity forms ``simultaneously'', it is 
convenient to label the hyperplane at which the singularity forms as
\(t=0\), and choose \(t\) to \emph{increase} towards the past (so that \(t\)
is positive in the region of interest). We can write down \(\mathcal{N}\) 
immediately in piecewise form:
\begin{equation} \begin{array}{ll}
   x^{\mu}(t,z,\phi) = (-t, t \cos \phi, t \sin \phi, z) & \mbox{cylinder} \\
   x^{\mu}(t,\theta,\phi) =
     (-t,t \sin \theta \cos \phi, t \sin \theta \sin \phi, t \cos \theta) &
     \mbox{right endcap} \\
   x^{\mu}(t,\theta',\phi) =
     (-t,t \sin \theta' \cos \phi, t \sin \theta' \sin \phi, t \cos \theta') &
     \mbox{left endcap} \\
   \end{array} \end{equation}
\begin{equation}
z \in [-\frac{L}{2},\frac{L}{2}]; \theta,\theta' \in [0,\frac{\pi}{2}];
\phi \in [0,2 \pi)
\end{equation}
The pieces are matched by identifying 
\(z=L/2\)
with \(\theta = \pi/2\), and \(z=-L/2\) with \(\theta' = \pi/2\).

The initial mass density is taken to be uniform over each component of
the collapsing shell. Because the shape is retained during collapse, the
density varies with time as 
\begin{equation}\sigma(t)=\left\{ \begin{array}{ll}
\frac{M}{2 \pi t L} & \mbox{cylinder} \\
\frac{m}{2 \pi t^{2}} & \mbox{endcaps}
\end{array} \right. \end{equation}
We will refer to \(M\) as 
``the total mass-energy of the cylindrical portion'', and \(m\) as ``the 
total mass-energy of each endcap''. Note that, except for one value of 
\(t\), the mass density is discontinuous where the cylinder meets the endcaps.

Again, we seek MTS's that have the same symmetries as \(\mathcal{N}\), since
the outermost MTS (which is the one of interest to us) must have these 
symmetries.
Because of the reflection-symmetry of the MTS about \(z=0\), we focus on 
\(z \geq 0\) and ignore the left endcap. Then, exploiting the axial symmetry,
we can define a candidate 2-surface by the equations
\begin{equation}t=\left\{ \begin{array}{ll}
f(z) & \mbox{cylinder} \\
V(\theta) & \mbox{endcap}
\end{array} \right. \end{equation}
It has been shown \cite{kpt:1992} that if \(f\) and \(V\) are to describe
an MTS, they must satisfy \cite{correction}

\begin{equation}
- 2 f \frac{d^{2}f}{dz^{2}} + 1 - \left( \frac{df}{dz} \right)^{2} = 
  \frac{8M}{L}
\label{MTS1}
\end{equation}
\begin{equation}
1 - \frac{1}{\sin \theta} \frac{d}{d\theta} \left( \sin \theta
  \frac{d}{d\theta} (\log V) \right) = \frac{4m}{V}
\label{MTS2}
\end{equation}
The MTS must be smooth at \(z=0\), regular at \(\theta=0\), but need only
be \(C^{1}\) at \(z=\frac{L}{2}\) since \(\mathcal{N}\) itself is only
\(C^{1}\) and the density is discontinuous there. Hence, we require
\begin{equation} \frac{df}{dz}(0)=0 \label{reflsym} \end{equation}
\begin{equation} V(\frac{\pi}{2})=f(\frac{L}{2}) \label{mc1} \end{equation}
\begin{equation}
\frac{d}{d\theta}(\log V)(\frac{\pi}{2})=-\frac{df}{dz}(\frac{L}{2})
\label{mc2}
\end{equation}
\begin{equation} \frac{dV}{d\theta}(0)=0 \label{rc} \end{equation}

The model is completely specified by the three parameters \(m\), \(M\) and 
\(L\). Since, as \(m\), \(M\) and \(L\) are scaled together, the solution will 
scale with them, it is useful to define \(\lambda=M/L\) and \(\mu=m/M\). Then 
a given pair \((\lambda,\mu)\) defines an equivalence class of solutions with
respect to overall change of mass/distance scale.

In the original presentation of the BIL model \cite{bil}, it was
demonstrated that for \(\lambda < 1/8\), ``the collapsing shell'' (meaning 
the intersection of \(\mathcal{N}\) with a flat hyperplane) is never a 
trapped or marginally-trapped surface, whereas for \(\lambda>1/8\), 
these intersection surfaces eventually become trapped. Our aim here is 
to investigate whether in the case \(\lambda < 1/8\) trapped surfaces 
exist on the shell that do not lie on flat 
hyperplanes. We therefore restrict attention to \(\lambda < 1/8\).

To investigate this issue, we proceed as follows. First, we fix \(M\) and
\(L\) and solve eqs.\ (\ref{MTS1}) and (\ref{reflsym}) on the interval
\([0,L/2]\), with \(t_{0} \equiv f(0)\) treated as an arbitrary free parameter.
(Note that the parameter \(m\) does not enter these equations.) A unique
solution to these equations exists \cite{kpt:1992} for all choices of
\(t_{0}>0\). The resulting value of \(f\) and its derivative
at \(z=L/2\) then generate the
required initial conditions (\ref{mc1}) and (\ref{mc2}) for the system
(\ref{MTS2}),(\ref{mc1}),(\ref{mc2}) and (\ref{rc}). This is an overdetermined 
system, so, in general, no solution will exist. However, in appendix A, we 
outline a proof (due to David Ross) that there always exists a unique choice 
of \(m\) such that a solution to these equations exists.

\begin{figure}
\centering
\includegraphics{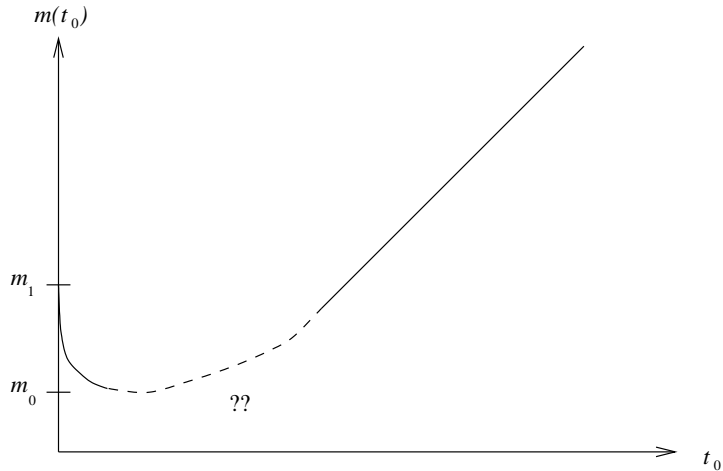}
\caption{Qualitative behavior of \(m\) vs. \(t_{0}\).}
\label{m}
\end{figure}

For each fixed \(M\) and \(L\), this unique solution defines a continuous
function \(m(t_{0})\). The asymptotic behavior of this function as \(t_{0}
\rightarrow 0\) and as \(t_{0} \rightarrow \infty\) is analyzed in appendix B.
As also noted there, it is not difficult to show that no solutions exist for 
\(m=0\).
This fact together with the continuity and asymptotic behavior of \(m(t_{0})\)
implies that there exists an \(m_{0}>0\) such that \(m(t_{0}) \geq m_{0}\)
for all \(t_{0}\). The resulting behavior of the function \(m(t_{0})\)
is sketched in fig. \ref{m}.

From inspection of fig. \ref{m}, we see that at each fixed \(M\) and \(L\),
there exist three qualitatively distinct regimes with regard to solutions
of the MTS equation. For \(0 \leq m < m_{0}\), there do not exist any MTS's
lying on \(\mathcal{N}\). On the other hand, for \(m_{0}<m<m_{1}\), there
exist at least two MTS's on \(\mathcal{N}\). Finally, for \(m\) sufficiently
large, there exists a unique MTS lying on \(\mathcal{N}\).

\begin{figure}
\centering
\includegraphics{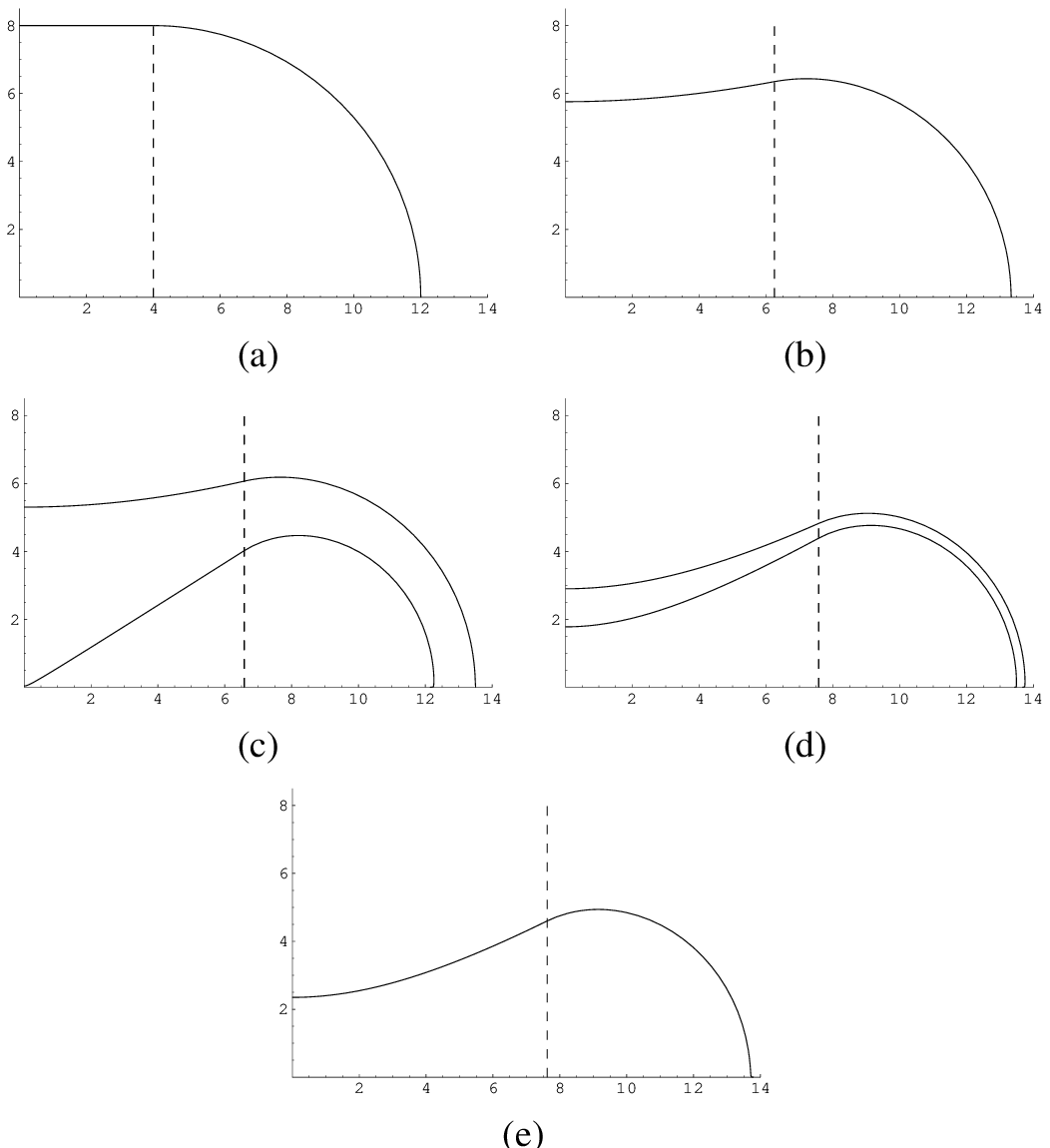}
\caption{
\(\mu\) held fixed (\(\mu=2\)), \(\lambda\) varied.
(a) \(\lambda=0.125\); (b) \(\lambda=0.08\);
(c) \(\lambda=\lambda_{1}=0.076\); (d) \(\lambda=0.066\);
(e) \(\lambda=\lambda_{0}=0.0656\)}
\label{mergeall}
\end{figure}

\begin{figure}
\centering
\includegraphics{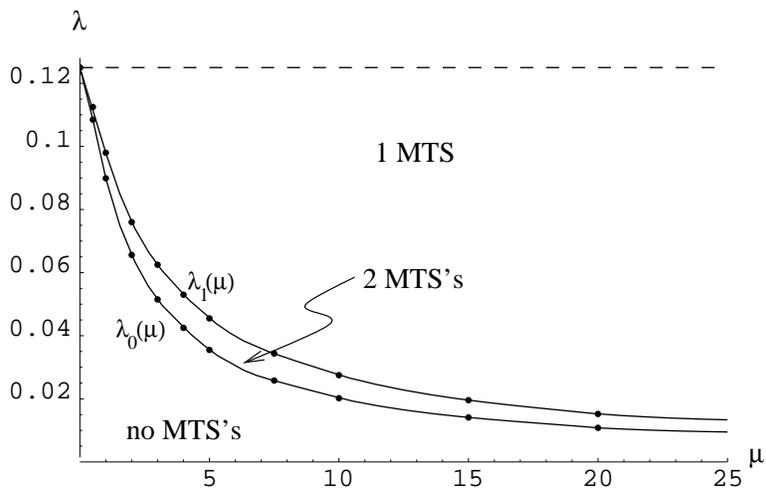}
\caption{Phase diagram, \(\lambda\) vs. \(\mu\)}
\label{phase}
\end{figure}

In order to gain more insight into the behavior of the MTS's, we have 
numerically
studied the solutions to the MTS equations for a wide range of the essential
parameters \(\lambda\) and \(\mu\).
\begin{figure}
\centering
\includegraphics{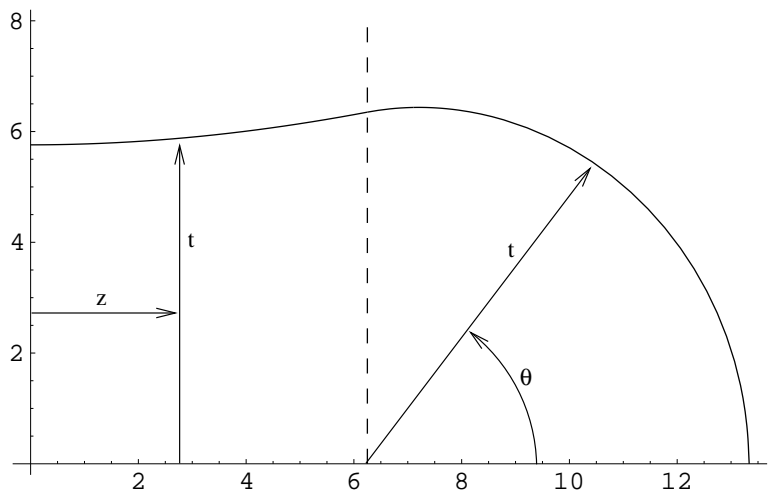}
\caption{Interpreting the plots. We represent the MTS on \(\mathcal{N}\)
by projecting it onto a flat hyperplane in the Minkowski region and then
suppressing the \(\phi\)-coordinate. Shown is a
typical MTS (\(\mu=2,\lambda=0.08\)). In the central region (to the left of
the dotted line), the ordinate yields the time to the past of
the singularity on null generator labelled by \(z\) on the cylindrical
portion of the surface. In the end region (to the right of the dotted line), 
the radius (measured from the intersection of the dotted line with
the abscissa) as a function of the polar angle \(\theta\) yields the 
time to the past of the singularity on null generator labelled by \(\theta\) 
on the hemispherical endcap.}
\label{sample}
\end{figure}
For any fixed \(\mu>0\), we find that when \(\lambda\) is sufficiently 
close to \(1/8\), there always exists a unique MTS (see fig. \ref{sample}). 
This MTS is the outer boundary of a trapped region. As \(\lambda\) is 
decreased holding \(\mu\) fixed (\emph{i.e.,\ }as the shell is 
made more prolate), one encounters a critical value, \(\lambda=\lambda_{1}\), 
beyond which a second MTS appears inside the first one,
almost touching the singularity. As \(\lambda\) is decreased further, 
the outer MTS and inner MTS migrate towards each other, until,
at a second critical value \(\lambda=\lambda_{0}\), the two surfaces merge and 
then disappear (see fig. \ref{mergeall}). For \(\lambda < \lambda_{0}\), no 
marginally-trapped
surfaces are found confined to the null hypersurface. These critical values
\(\lambda_{0}\) and \(\lambda_{1}\) describe curves in the \((\lambda,\mu)\)
plane, dividing parameter space into regions for which qualitatively
different types of solutions occur (fig. \ref{phase}).

\begin{figure}
\centering
\includegraphics{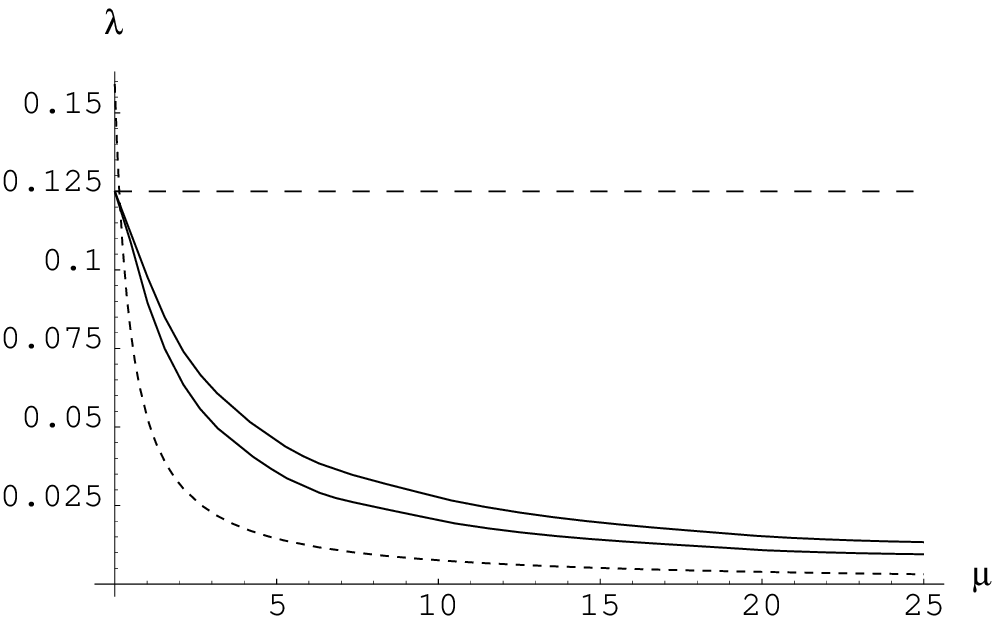}
\caption{The same phase diagram as in fig. \ref{phase}, with the
function \(\lambda = (2 \pi (1+2 \mu))^{-1}\) also plotted (dashed line).}
\label{phase2}
\end{figure}

The disappearance of trapped surfaces on \(\mathcal{N}\) when \(\lambda
<\lambda_{0}(\mu)\) is in accord with the qualitative behavior one might
have expected from the hoop conjecture. 
To investigate this more quantitatively, we interpret the ``mass''
appearing in the general statement of the hoop conjecture (see section 1)
as being the total mass-energy, \(M_{\mathrm{tot}}=M+2m\), of the shell,
and we take the ``circumference'' of a 2-surface on \(\mathcal{N}\) to
be the pole-to-pole circumference \(2L+4I\), where
\begin{equation}
I= \int_{0}^{\pi/2} V(\theta) d\theta
\label{I}
\end{equation}
Since \(C \geq 2L\) for all 2-surfaces on \(\mathcal{N}\), with equality
holding to arbitrarily good accuracy as one approaches the singularity,
if this interpretation of the hoop conjecture is correct, one would
expect that trapped surfaces will exist on \(\mathcal{N}\) if and only
if \(2L \leq 4 \pi (M+2m)\), \emph{i.e.\ }if and only if
\begin{equation}
\frac{1}{\lambda} \leq 2 \pi (1+2 \mu)
\label{hoop}
\end{equation}
Thus, the hoop conjecture suggests that the curve defined by equality in
eq.\ (\ref{hoop}) should be the ``dividing line'' between the regimes
where trapped surfaces do and do not occur on \(\mathcal{N}\). This
curve is plotted as a dashed line in fig. \ref{phase2}. It can be seen
from the figure that this line lies remarkably close to the actual
dividing line\footnote{In comparison, the ``isoperimetric inequality'' 
\(A \leq 16 \pi M_{\mathrm{tot}}^{2}\) is satisfied by a wide margin for 
the MTS's 
occurring in the highly prolate case. Thus, the validity of this inequality 
\cite{gg:1997} does not provide a ``reason'' why MTS's do not exist for 
\(\lambda<\lambda_{0}(\mu)\).} \(\lambda = \lambda_{0}(\mu)\).

Furthermore, even better quantitative agreement with expectations from
the hoop conjecture is obtained if we consider the ratio \(C/M_{\mathrm{tot}}\)
for the ``last'' MTS, \emph{i.e.\ }if we examine, as a function of \(\mu\),
the quantity
\begin{equation}
\frac{C}{M_{\mathrm{tot}}} = \frac{2L+4I}{M+2m} = \frac{2+\frac{4I}{L}}
  {(1+2 \mu) \cdot \lambda_{0}(\mu)}
\end{equation}
where \(I\) is given by eq.\ (\ref{I}) above. In numerical solutions, we find 
that the ratio \(C/M_{\mathrm{tot}}\) decreases monotonically with \(\mu\).
Although it is as much as 30\% greater than the \(4 \pi\) appearing in the 
statement of the hoop conjecture in the limit \(\mu \rightarrow 0\), it 
is only about 2\% less than \(4 \pi\) in the limit \(\mu \rightarrow \infty\) 
(the limit of extreme prolateness).

Does the absence of trapped surfaces on \(\mathcal{N}\) when \(\lambda<
\lambda_{0}(\mu)\) suggest that trapped surfaces are completely absent from
the spacetime? If so, this could be viewed as evidence that the singularity
produced by the collapse of the shell is naked in this regime. However, we
do not believe that the absence of trapped surfaces on \(\mathcal{N}\) when
\(\lambda<\lambda_{0}(\mu)\) provides strong evidence for the absence of
trapped surfaces in the spacetime. When \(\lambda=1/8\) and \(\mu>0\), the
outermost MTS is the intersection of \(\mathcal{N}\) with a hyperplane at
a time \(t=t_{\mathrm{MTS}}\). The intersection of \(\mathcal{N}\) with any
hyperplane with \(0<t<t_{\mathrm{MTS}}\) is a trapped surface \cite{bil},
so there is a large ``trapped region'' on \(\mathcal{N}\). There is also
a large trapped region to the future of \(\mathcal{N}\). Namely, if
\(\mathcal{T}\) denotes a trapped surface on \(\mathcal{N}\) and
\(\mathcal{L}\) denotes the null hypersurface generated by the outgoing
null geodesics from \(\mathcal{T}\), then (by the Raychaudhuri equation)
every cross-section of \(\mathcal{L}\) also is trapped. As \(\lambda\) is
decreased (with \(\mu\) held fixed) the trapped region on \(\mathcal{N}\)
begins to ``migrate away'' from the singularity at \(t=0\) near the poles
(\(\theta,\theta'=0\)).
At \(\lambda=\lambda_{1}(\mu)\), the trapped region on \(\mathcal{N}\) begins
to migrate away from the singularity near the equator (\(z=0\)) as well.
For \(\lambda_{0}(\mu)<\lambda<\lambda_{1}(\mu)\), the trapped region on 
\(\mathcal{N}\)
presumably is the region bounded by the two MTS's present on \(\mathcal{N}\)
(see fig. \ref{mergeall}d). However, trapped surfaces also must continue
to exist in the region to the future of \(\mathcal{N}\) swept out by outgoing
null geodesics orthogonal to the trapped surfaces on \(\mathcal{N}\). When
\(\lambda=\lambda_{0}\), there is a single MTS, \(\mathcal{M}\), on
\(\mathcal{N}\), but no trapped surfaces on \(\mathcal{N}\). However, since
the shear of the outgoing null geodesics orthogonal to \(\mathcal{M}\) is
nonvanishing, there must continue to exist trapped surfaces to the future
of \(\mathcal{N}\). Thus, the disappearance of trapped surfaces on
\(\mathcal{N}\) at \(\lambda=\lambda_{0}\) does not signal the disappearance
of trapped surfaces from the spacetime, but rather their migration to the
future of \(\mathcal{N}\). However, the issue of whether or not trapped 
surfaces continue to be present as \(\lambda \rightarrow 0\) cannot be
investigated until we have a much more detailed knowledge of the spacetime 
exterior to \(\mathcal{N}\).

\begin{center}
\large \textbf{Acknowledgements} \normalsize
\end{center}

We wish to thank David Ross for providing us with the existence and
uniqueness proofs sketched in Appendix A. This research was supported
in part by NSF grant PHY 95-14726 to the University of Chicago. Two of
us (K.P.T. and R.M.W.) also wish to thank the Erwin Schr\"{o}dinger
Institute for hospitality during the summer of 1994, where some of the
ideas for this research project arose.

\appendix

\section{Existence and uniqueness of solutions}

In this appendix, we outline a proof that in the BIL model, given any 
\(t_{0}>0\), 
there exists a unique value of \(m\) yielding an MTS on \(\mathcal{N}\) 
with \(f(0)=t_{0}\). (That is, we prove that the function \(m(t_{0})\) 
described in the text is well-defined for all \(t_{0}>0\).) 
The key ideas in the existence and uniqueness proof are due to David Ross.

We fix \(M,L,\) and \(t_{0}\), and leave \(m\) to be determined. We wish
to find a solution of the following system of equations:

\begin{equation}
- 2 f \frac{d^{2}f}{dz^{2}} + 1 - \left( \frac{df}{dz} \right)^{2} = 
  \frac{8M}{L}
\label{f1}
\end{equation}
\begin{equation} \frac{df}{dz}(0)=0 \label{f2} \end{equation}
\begin{equation} f(0)=t_{0} \label{f3} \end{equation}
\begin{equation}
1 - \frac{1}{\sin \theta} \frac{d}{d\theta} \left( \sin \theta
  \frac{d}{d\theta} (\log V) \right) = \frac{4m}{V}
\label{V1}
\end{equation}
\begin{equation} V(\frac{\pi}{2})=f(\frac{L}{2}) \label{V2} \end{equation}
\begin{equation}
\frac{d}{d\theta}(\log V)(\frac{\pi}{2})=-\frac{df}{dz}(\frac{L}{2})
\label{V3}
\end{equation}
\begin{equation} \frac{dV}{d\theta}(0)=0 \label{V4} \end{equation}

One can derive the following implicit solution \cite{kpt:1992} for the first 
three equations (\ref{f1}-\ref{f3})
\begin{equation}
\frac{2 \alpha z}{t_{0}} = \frac{1}{w-1} + \frac{1}{w+1}
  - \log \left( \frac{w-1}{w+1} \right)
\label{implicit}
\end{equation}
\begin{equation}
w = \sqrt{\frac{f}{f-t_{0}}}
\label{implicit2}
\end{equation}
where
\begin{equation}
\alpha = \sqrt{1-\frac{8M}{L}}
\end{equation}
This solution generates certain values \(f(L/2)>0\) and \(0<df/dz(L/2)<1\); 
taking these values as given (determined by our
choices of \(M,L,\) and \(t_{0}\)), we focus on (\ref{V1}-\ref{V4}). 
This system of equations, a 2nd-order ODE with three boundary conditions, 
would be overdetermined if we also specified \(m\). However, we instead
recast eqs. (\ref{V1}), (\ref{V3}), and (\ref{V4}) in terms of
the ``scale-invariant'' function \(\hat{V}=V/m\), which obeys
\begin{equation}
1 - \frac{1}{\sin \theta} \frac{d}{d\theta} \left( \sin \theta
  \frac{d}{d\theta} (\log \hat{V}) \right) = \frac{4}{\hat{V}}
\label{vsi1}
\end{equation}
\begin{equation}
\frac{d}{d\theta}(\log \hat{V})(\frac{\pi}{2})=-\frac{df}{dz}(\frac{L}{2})
\label{vsi2}
\end{equation}
\begin{equation} \frac{d\hat{V}}{d\theta}(0)=0 \label{vsi3} \end{equation}
These equations do not involve \(m\).
The remaining boundary condition (\ref{V2}) is then satisfied if and only
if \(m\) is chosen to be
\begin{equation}
m=\frac{f(L/2)}{\hat{V}(\frac{\pi}{2})}.
\label{determinationofm}
\end{equation}
It follows that if there exists a unique solution to (\ref{vsi1}-\ref{vsi3}),
then there exists a unique value of \(m\) yielding an MTS with the 
chosen \(t_{0}\).

Now write \(k=-df/dz(L/2)\), let primes denote derivatives 
with respect to \(\theta\), and define \(W \equiv (\log \hat{V})'\). This 
gives
\begin{equation}
1 - \frac{1}{\sin \theta} ( W \sin \theta)' = \frac{4}{\hat{V}}
\label{kefir1}
\end{equation}
\begin{equation}
\hat{V}'=W\hat{V}
\label{kefir2}
\end{equation}
\begin{equation}
W(\frac{\pi}{2})=k
\label{kefir3}
\end{equation}
\begin{equation}
W(0)=0
\label{kefir4}
\end{equation}
In order to prove the existence and uniqueness of solutions to this
boundary-value problem, we convert it to an initial-value problem
and use a ``shooting argument'', \emph{i.e.\ }we consider solutions to
(\ref{kefir1}), (\ref{kefir2}), (\ref{kefir4}) and
\begin{equation}
\hat{V}(0)=\hat{V}_{0}>0
\label{kefir5}
\end{equation}
and then demonstrate that one can choose \(\hat{V}_{0}\) uniquely in order to
satisfy the boundary condition (\ref{kefir3}). We will first
prove local existence and uniqueness of solutions to (\ref{kefir1}),
(\ref{kefir2}), (\ref{kefir4}) and (\ref{kefir5}) around \(\theta=0\),
then show global existence and uniqueness on the interval \([0,\pi/2]\)
for at least some choices of \(\hat{V}_{0}\), and finally argue that 
(\ref{kefir3}) can always be satisfied.

To begin, we note that (\ref{kefir1}) and (\ref{kefir4}) are equivalent
to the integral equation
\begin{equation}
W=\frac{1}{\sin \theta} \int_{0}^{\theta} \left(1-\frac{4}{\hat{V}(t)}
  \right) \sin t~dt
\label{ivp1}
\end{equation}
whereas eqs. (\ref{kefir2}) and (\ref{kefir5}) are equivalent to
\begin{equation}
\hat{V}(\theta) = \hat{V}_{0} + \int_{0}^{\theta} W(t) 
    \hat{V}(t)~dt
\label{ivp2}
\end{equation}
To establish local existence of solutions to (\ref{ivp1}) and (\ref{ivp2}), 
we use a variant of ``Picard iteration''. Consider the iteration scheme
given by
\begin{equation}
W_{j}(\theta) = \frac{1}{\sin \theta} \int_{0}^{\theta} \left(1-
  \frac{4}{\hat{V}_{j-1}(t)} \right) \sin t~dt
\end{equation}
\begin{equation}
\hat{V}_{j}(\theta) = \hat{V}_{0} + \int_{0}^{\theta} W_{j-1}(t) 
    \hat{V}_{j-1}(t)~dt
\end{equation}
with initial ``guesses'' \(\hat{V}_{j=1}=\hat{V}_{0}\) and \(W_{j=1}=0\). It
can be shown, by demonstrating that the \(W_{j}\)'s and \(\hat{V_{j}}\)'s are
bounded in some small interval beginning at \(\theta=0\), that the iterates 
converge uniformly as
\begin{equation}
|\hat{V}_{j}-\hat{V}_{j-1}| < \frac{(C \theta)^{j}}{j!}
\end{equation}
\begin{equation}
|W_{j}-W_{j-1}| < \frac{(C \theta)^{j}}{j!}
\end{equation}
where \(C\) is a constant.
Because the convergence is uniform, the limit functions 
\(\hat{V}=\lim \hat{V}_{j}\) and \(W=\lim W_{j}\) satisfy
(\ref{ivp1}) and (\ref{ivp2}), and so constitute a (local) solution to the 
initial-value problem.

Local uniqueness is proved similarly.  Let \(\hat{V}_{1},W_{1}\)
and \(\hat{V}_{2},W_{2}\) both be solutions of the initial-value
problem. Then, again using the fact that both of these solutions
are bounded on a sufficiently small interval starting at
\(\theta=0\), one can show that for all positive integers \(j\), 
\begin{equation}
|\hat{V}_{1}-\hat{V}_{2}| < P \frac{(2Q)^{j} \theta^{j}}{j!}
\end{equation}
\begin{equation}
|\hat{W}_{1}-\hat{W}_{2}| < P \frac{(2Q)^{j} \theta^{j}}{j!}
\end{equation}
where \(P\) and \(Q\) are constants. Therefore, \(V_{1}=V_{2}\)
and \(W_{1}=W_{2}\), and solutions of the initial-value problem
are unique near \(\theta=0\).

By a standard theorem in the theory of ODE's \cite{result}, then, this unique
local solution to eqs. (\ref{ivp1}) and (\ref{ivp2}) can be extended away 
from \(\theta=0\) until it becomes unbounded.
It remains to be shown that \(\hat{V}_{0}\) can be 
chosen so that the solution can be extended all the way to \(\theta=\pi/2\)
and satisfies \(W(\frac{\pi}{2})=k\). It is not difficult to prove that, for 
fixed \(\theta>0\), \(\hat{V}\) and \(W\) are strictly increasing functions 
of \(\hat{V}_{0}\), so if we can
find a \(\hat{V}_{0}\) which yields the correct matching condition, this
\(\hat{V}_{0}\) will be unique.

The analysis of global existence of solutions on the interval \([0,\pi/2]\)
naturally divides into the following three cases.
\begin{itemize}
\item \(\hat{V}_{0}>4\). Then \(\hat{V}\) and \(W\) are strictly increasing
functions of \(\theta\). Furthermore, \(\hat{V}\) and \(W\) are bounded
on the entire interval \([0,\pi/2]\), according to
\begin{equation}
\hat{V}_{0} \leq \hat{V}(\theta) \leq \hat{V}_{0} \exp 
  \left( \int_{0}^{\theta} \frac{1-\cos t}{\sin t}~dt \right)
\end{equation}
\begin{equation}
0 \leq W(\theta) < \frac{1-\cos \theta}{\sin \theta}
\end{equation}
with equalities occurring only at \(\theta=0\). As \(\hat{V}_{0} \rightarrow
\infty\), \(W(\pi/2) \rightarrow 1\), while as \(\hat{V}_{0} \rightarrow 4\),
\(W(\pi/2) \rightarrow 0\). By continuous dependence of solutions on initial
data, it follows that all values of \(W(\pi/2)\) satisfying \(0<W(\pi/2)<1\)
can be achieved.
\item \(\hat{V}_{0}=4\). Then \(\hat{V}(\theta)=4\) and \(W(\theta)=0\).
\item \(\hat{V}_{0}<4\). Then \(W<0\) for \(\theta>0\), and, for
fixed \(\hat{V}_{0}\), \(\hat{V}(\hat{V}_{0},\theta)\) is a strictly 
decreasing function of \(\theta\). (Recall that for fixed \(\theta\),
\(\hat{V}(\hat{V}_{0},\theta)\) is also a strictly increasing function
of \(\hat{V}_{0}\).) In this case, it is possible that for some 
\(\hat{V}_{0}\), the solution ``blows up'' (\(\hat{V} \rightarrow 0\) 
and \(W \rightarrow - \infty\)) before reaching \(\theta=\pi/2\), in which case
a global solution does not exist. (It is not difficult to show that this
is the only way in which
a global solution can fail to exist.) We wish to show that given any
\(k \in (-\infty,0)\), \(\hat{V}_{0}\) can nevertheless be chosen to 
yield a global solution with \(W(\pi/2)=k\). First, we note that
since \(\hat{V}=4\) is a global solution, it follows from continuous
dependence of solutions on initial data that there exist global solutions on 
the
interval \([0,\pi/2]\) for some \(\hat{V}_{0}<4\). Let \(c\) be the greatest
lower bound of the set of \(\hat{V}_{0}\) for which global solutions exist.
Then, from the monotonicity properties of \(\hat{V}\), it follows that global
solutions exist for all \(\hat{V}_{0} \in (c,4)\).
Let \(p\) be the greatest lower bound of \(\hat{V}(\pi/2)\) for these
global solutions (so that, by monotonicity, as \(\hat{V}_{0} \rightarrow c\),
\(\hat{V}(\hat{V}_{0},\pi/2) \rightarrow p\)). Then there exist three
possibilities: (i) If \(p=0\), then it can be shown that as 
\(\hat{V}(\hat{V}_{0},\pi/2) \rightarrow 0\), \(W(\pi/2) \rightarrow -\infty\),
Therefore, by continuous dependence of solutions on initial data, any 
value \(W(\pi/2) \in (-\infty,0)\) can be obtained by some choice of 
\(\hat{V}_{0} \in (c,4)\). (ii) If \(p>0\) and \(c=0\), then we contradict
the fact that \(\hat{V}(\theta)\) is strictly decreasing in \(\theta\). 
(iii) If \(p>0\) and \(c>0\), then one can show, using continuous dependence 
of solutions on initial
conditions and the fact that solutions can fail to exist globally only by
\(\hat{V}\) approaching 0 at some \(\theta \leq \pi/2\), that there exists
a global solution with \(\hat{V}_{0}=c\) and \(\hat{V}(\pi/2)=p\). 
However, again using continuous dependence of solutions on initial data,
it then follows that there exist global solutions with 
\(\hat{V}_{0}<c\). Thus we obtain a contradiction. Consequently, only
case (i) is possible.

\end{itemize}

In summary, for any \(\hat{V}_{0}>c\), we can obtain a global solution to the
initial-value problem on the interval 
\([0,\pi/2]\), and we can uniquely choose \(\hat{V}_{0}\) in this
domain in order to achieve any \(W(\pi/2)<1\).  (Recall that we only needed 
\(-1<W(\pi/2)<0\).) Therefore, there exists a unique solution to the 
boundary-value problem (\ref{vsi1}-\ref{vsi3}), and consequently a unique 
value of \(m\), determined by (\ref{determinationofm}), yielding an MTS with 
the chosen \(t_{0}\).

\section{Limiting behavior of \(m(t_{0})\)}

Given \(M\) and \(L\), we showed in appendix A that \(m\) is determined 
by a choice of \(t_{0}\). This dependence is mediated by the 
``matching conditions'' between \(f\) and \(V\), the center and end solutions.
We write
\begin{equation}
X(t_{0})=f(t_{0},z=L/2)
\end{equation}
\begin{equation}
Y(t_{0})=\frac{\partial f}{\partial z}(t_{0},z=L/2)
\end{equation}
where \(f(t_{0},z)\) denotes the center solution, \(f(z)\) (see eqs.
(\ref{implicit}),(\ref{implicit2})), with 
\(f(0)=t_{0}\). It is convenient to view \(X\) and \(Y\) as independent
variables and to view \(m\) as a function of \(X\) and \(Y\),
\begin{equation}
m=m(X,Y)
\end{equation}
We may then write
\begin{equation}
\frac{dm}{dt_{0}}=\frac{\partial m}{\partial X} \frac{dX}{dt_{0}}
   + \frac{\partial m}{\partial Y} \frac{dY}{dt_{0}}
\label{dmdt}
\end{equation}
In this appendix, we derive the limiting behavior as \(t_{0} \rightarrow 0\)
and as \(t_{0} \rightarrow \infty\) of each of the terms
on the RHS of this equation, and so discover the limiting behavior of 
\(m(t_{0})\).

We first examine how \(m\) changes with \(X\) when \(Y\)
is held fixed. Since \(m\) is determined by solving the system of equations
(\ref{vsi1}-\ref{vsi3}), and these equations are independent of
\(X\), it follows that if \(X\) is changed, keeping \(Y\) fixed, we obtain 
the same solution \(\hat{V}\). From eq.\ (\ref{determinationofm}), we
therefore obtain
\begin{equation}
\frac{\partial m}{\partial X} =\frac{1}{\hat{V}(\pi/2)}
\end{equation}

Next, we change \(Y\) while holding \(X\) fixed. Since we
found in the previous appendix that \(\hat{V}(\pi/2)\) and \(W(\pi/2)\) 
are strictly monotonically increasing functions of \(\hat{V}_{0}\),
an increase in \(Y\) results in a decrease in \(W(\pi/2)\), a
decrease in \(\hat{V}_{0}\), and a decrease in \(\hat{V}(\pi/2)\).
By eq.\ (\ref{determinationofm}), the required \(m\) must
increase, and so
\begin{equation}
\frac{\partial m}{\partial Y}>0
\label{dmdf'}
\end{equation}

For the remaining terms, we need to refer to the implicit solution 
(\ref{implicit}) and (\ref{implicit2}) for \(f(z)\), as well as the first 
integral of eq.\ (\ref{f1})
\begin{equation}
\frac{\partial f}{\partial z}(t_{0},z) = \alpha \sqrt{1-\frac{t_{0}}
  {f(t_{0},z)}}
\label{firstintegral}
\end{equation}
By taking the derivatives with respect to \(t_{0}\) of the implicit solution
(\ref{implicit}) and the first integral (\ref{firstintegral}) we obtain,
respectively,
\begin{equation}
\frac{dX}{dt_{0}}= \frac{X}{t_{0}} -
  \frac{\alpha L}{2 t_{0}} \sqrt{1-\frac{t_{0}}{X}}
\label{dxdt0}
\end{equation}
\begin{equation}
\frac{dY}{dt_{0}} = - \frac{\alpha^{2} L}{4 X^{2}}
\label{dydt0}
\end{equation}

Although our knowledge of \(dm/dt_{0}\) is limited by the fact that we
have only
qualitative knowledge of \(\partial m/\partial Y\), eq.\ (\ref{dmdf'}),
we can nevertheless derive its exact behavior in the limits
\(t_{0} \rightarrow 0\) and \(t_{0} \rightarrow \infty\).
When \(t_{0} \ll L\), the implicit solution for \(f\) yields,
to lowest order,
\begin{equation}
X \approx \frac{\alpha L}{2} + \frac{t_{0}}{2} \log
   \left( \frac{t_{0}}{\alpha L} \right)
\end{equation}
implying
\begin{equation}
\frac{dX}{dt_{0}} \sim \log \left(
  \frac{t_{0}}{\alpha L} \right) \rightarrow - \infty~\mbox{as}~
  t_{0} \rightarrow 0
\end{equation}
On the other hand, since \(0<\partial f/\partial z(t_{0},z)<\alpha\), 
\begin{equation}
t_{0}<X<t_{0}+\alpha L/2
\end{equation}
Then eq.\ (\ref{dxdt0}) gives
\begin{equation}
\frac{dX}{dt_{0}} \rightarrow 1~\mbox{as}~
  t_{0} \rightarrow \infty
\end{equation}

Since, by the above, \(X \rightarrow \alpha L/2\) as \(t_{0} \rightarrow 0\)
and \(X \rightarrow \infty\) as \(t_{0} \rightarrow \infty\),
we have from eq.\ (\ref{dydt0})
\begin{equation}
\frac{dY}{dt_{0}}
  \rightarrow
  \left\{ \begin{array}{cl}
   - \frac{1}{L} & \mbox{as}~t_{0} \rightarrow 0 \\
   0 & \mbox{as}~t_{0} \rightarrow \infty
  \end{array} \right.
\end{equation}

According to eq.\ (\ref{firstintegral}),
\begin{equation}
Y \rightarrow 
  \left\{ \begin{array}{cl}
   \alpha & \mbox{as}~t_{0} \rightarrow 0 \\
   0 & \mbox{as}~t_{0} \rightarrow \infty
  \end{array} \right.
\end{equation}
Now, applying a result of the previous appendix, \(\hat{V}(\pi/2)\) 
approaches some value between 0 and 4 in the first limit, and 
\(\hat{V}(\pi/2)\) goes to 4 in the second limit.
Thus
\begin{equation}
\frac{\partial m}{\partial X} \rightarrow
  \left\{ \begin{array}{cl}
   \mbox{const.}~>\frac{1}{4} & \mbox{as}~t_{0} \rightarrow 0 \\
   \frac{1}{4} & \mbox{as}~t_{0} \rightarrow \infty
  \end{array} \right.
\end{equation}
Returning to eq.\ (\ref{dmdt}), we have that
\begin{equation}
\frac{dm}{dt_{0}} \rightarrow - \infty~\mbox{as}~t_{0} \rightarrow 0
\end{equation}
whereas \(m\) approaches a finite limiting value, \(m_{1}\), as \(t_{0}
\rightarrow 0\)
(since there are no special difficulties encountered in solving 
(\ref{vsi1}-\ref{vsi3}) when \(\partial f/\partial z(L/2)=\alpha\)).
In the other limit, we have simply
\begin{equation}
\frac{dm}{dt_{0}} \rightarrow \frac{1}{4}~\mbox{as}~t_{0} \rightarrow 
\infty
\end{equation}

Finally, note that when \(m=0\), we can explicitly solve
for \(V(\theta)\), finding that \((\log V)'(\pi/2)=1\). However, it
is impossible to choose \(t_{0}\) in order to satisfy the corresponding
matching condition \(\partial f/\partial z(L/2)=-1\). Thus there exists no 
MTS confined to the null hypersurface \(\mathcal{N}\) when \(m=0\).

The above results imply that \(m(t_{0})\) has the behavior described
in section 4 (see figure \ref{m} above).

\end{document}